\documentstyle[floats,aps,psfig,twocolumn]{revtex}
\begin{document}
\title{Charge ordering and spin- Peierls transition in 
\mbox{$\alpha$'-NaV$_2$O$_5$}}
\author{P. Thalmeier}
\address{Max-Planck-Institut f\"ur Chemische Physik fester Stoffe,
D-01187 Dresden, Germany}
\author{P. Fulde}
\address{Max-Planck-Institut f\"ur Physik komplexer Systeme,
D-01187 Dresden, Germany}
\date {\today}
\maketitle
\begin{abstract}
Recent X-ray analysis and NMR experiments have shown that $\alpha$'-NaV$_2$O$_5$ is not a conventional spin-Peierls (SP) compound because above the transition at 33K it is in a homogeneous mixed valent (MV) state and atomic spin chains do not exist. Furthermore, thermal expansion experiments have identified the existence of a double phase transition. We present a theoretical model which qualitatively explains these observations. We derive an effective Hamiltonian for the insulating state which describes the transitions as primary charge order leading to linear spin chains in b- direction and a secondary SP- transition. We show that this scenario explains in a natural way the anomalous BCS- ratio observed in $\alpha$'-NaV$_2$O$_5$.  
\end{abstract}
\vspace{0.5cm}
\pacs{PACS 75.10.Jm\\
thalm@cpfs.mpg.de\\
Fax: ++49-351-871-1199\\
Tel: ++49-351-871-1117\\[1cm]
postal address:\\
c/o Max-Planck-Institut PKS\\
N\"othnitzer Str. 38\\
D-01187 Dresden}



The discovery of the first inorganic spin-Peierls (SP) compound CuGeO$_3$ \cite{Hase} has renewed interest in a phase transition originally observed in a few organic compounds \cite{Jacobs}. The spin- chain dimerization in CuGeO$_3$ below T$_{SP}$= 14K was investigated in detail using neutron and Raman scattering and also by thermal expansion and susceptibility measurements. Theoretical understanding of the SP- transition as achieved by Pytte \cite{Pytte}. In case the interchain coupling is sufficiently large a mean field HF approximation is possible and then the BCS ratio of 2$\Delta$(0)/kT$_{SP}$=3.52 for the dimerization gap $\Delta$(0) is obtained. This is close to the value observed in CuGeO$_3$ and in the organic SP- compounds.\\
Later on another inorganic 3d- insulator, $\alpha$'-NaV$_2$O$_5$ was investigated \cite{Isobe}. It can be viewed as a layered structure with corrugated planes composed of 3d-V bonds oriented along the segments of zig-zag chains and along the 'rungs' parallel to the a-axis which connect different zig-zag chains (Fig.1a). From the temperature dependence of the susceptibility the existence of 1D spin- chains and the opening of an isotropic spin gap below 33K has been concluded \cite{Isobe}. According to the original X-ray structure analysis \cite{Carpy} of $\alpha$'-NaV$_2$O$_5$ vanadium atoms exist in V$^{4+}$(S=$\frac{1}{2}$) as well as V$^{5+}$(S=0) ionization states. The V$^{4+}$- species then forms linear AF spin chains along the b- axis which has been assumed to be the case at {\em all} temperatures. Therefore $\alpha$'-NaV$_2$O$_5$ was interpreted as another example of an inorganic SP- system which undergoes dimerization of the V$^{4+}$- spin chains below 33K.\\
Recently it has become clear that this simple view is not adequate. Firstly, the BCS- ratio has the anomalously high value of 6.44 in  $\alpha$'-NaV$_2$O$_5$. Furthermore, an important observation was made in thermal expansion measurements \cite{Koeppen}: The phase transition in this compound consists of {\em two} very close transitions at T$_{c1}$=33K and T$_{c2}$=32.7K with first order and continuous character, respectively. The latter transition is connected with the appearance of a large spontaneous strain. This observation initiated a new X-ray structure analysis \cite{Chatt1} which had a very surprising outcome. Contrary to the original analysis \cite{Carpy} a centrosymmetric space group (Pmmn) with {\em only one} crystallographic vanadium position was found to give the best structure refinement at room temperature \cite{Chatt2}. This implies that $\alpha$'-NaV$_2$O$_5$ is in a homogeneous insulating MV state with formal valence n=4.5. Therefore it cannot be viewed as a system of uniform atomic spin chains of  V$^{4+}$(S=$\frac{1}{2}$) ions above T$_{c1}$. In addition NMR experiments \cite{Ohama} have confirmed the high temperature homogeneous MV state and in addition have proven that below the (double) phase transition two inequivalent V-sites evolve.\\
These findings lead us to the following conjecture: The primary order parameter (OP) in $\alpha$'-NaV$_2$O$_5$ is a charge order of  V$^{4+}$/V$^{5+}$ below T$_{c1}$ which immediately leads to the possibility of a secondary order parameter below T$_{c2}$ connected with the evolution of a spin gap and a large lattice distortion. Since the latter is characterized by a wave vector Q= ($\frac{1}{2}$, $\frac{1}{2}$, $\frac{1}{4}$,) (units of $2\pi$/a  etc.) a dimerization along the b axis is present and therefore it is possible that the secondary OP is mainly of the SP type along the b- chain direction. Other possibilities \cite{Chatt1,Seo} cannot be excluded at present since the precise nature of  V$^{4+}$/V$^{5+}$ charge ordering below  T$_{c1}$ is not yet known. However inelastic neutron scattering results \cite{Yoshi} seem to favor a charge order with 1D character along b. Irrespective of the details a theoretical model for $\alpha$'-NaV$_2$O$_5$ must address the following facts:\\
(1) at high temperature one has a homogeneous insulating MV state.
(2) there are two close consecutive phase transitions with the primary order parameter driven by  V$^{4+}$/V$^{5+}$ charge ordering. 
(3) the low- temperature phase shows a spin gap with anomalous high BCS- ratio.
(4) there is a lattice distortion including a doubling of the unit cell along b- direction.\\[0.5cm]
In this letter we present and investigate such a model. First we discuss basic features of the electronic structure. LDA- calculations \cite{Smol} suggest that one has vanadium d-bands with a dispersion of $\simeq$ 0.5 eV caused by effective hopping (t,t' along the zig-zag chain) and a bonding-antibonding gap of 2$\tilde{t}$ $\simeq$ 1eV where $\tilde{t}$ is the effective d-d hybridization (Fig.1a) via oxygen p-orbitals in the rungs between zig-zag chains. Because there is one vanadium-3d electron per V-V rung the Fermi level lies in the middle of the bonding bands, which means that LDA predicts a metallic state. The origin of the insulating state of $\alpha$'-NaV$_2$O$_5$ is therefore not obvious. In \cite{Smol,Mack} it was assumed that an on-site Coulomb repulsion U is sufficient to achieve an insulator. In fact if 2t$<$U$<$2$\tilde{t}$ the (half filled) bonding band may acquire a Mott Hubbard correlation gap. However in the regime 2t$<$2$\tilde{t}<$U$<\infty$ relevant for $\alpha$'-NaV$_2$O$_5$ it is not clear whether an on-site U is sufficient to obtain an insulating state or if in addition inter-site interactions are needed.\\ 
The essential reason for our introduction of the intersite repulsion $\tilde{V}$,V,V' (Fig.1) however is a different one: They are necessary to obtain the observed charge ordering transition irrespective of the nature of the insulating ground state. In fact a Hartree Fock calculation \cite{Seo} shows that U$>\tilde{V}$,V,V'$>\tilde{t}$,t,t' must hold in order to achieve complete charge order and moment formation at zero temperature. This is also the limit which we consider here. In this case we take the interaction terms as unperturbed Hamiltonian H$_0$. Its basis states $|$n$_{1\sigma}$,n$_{2\sigma '}\rangle$ are defined by the occupation numbers n$_{l\sigma}$ where l=1,2 denotes one of the V- atoms in each rung and $\sigma,\sigma'$ denotes the spin. On the average each rung accomodates one d- electron and for an insulating state, irrespective of its origin  only virtual hopping of electrons between rungs is possible. Therefore one can restrict to the one-particle (1-p) subspace for each rung which makes it convenient to introduce an Ising pseudospin variable $\tau_z$=n$_2$-n$_1$ with values $\pm$1 according to whether the 3d electron occupies the left (l=1) or right (l=2) V atom of each rung. In addition one has the real 3d spin $\frac{1}{2}\vec{\sigma}_l$. In this subspace H$_0$ reduces to

\begin{eqnarray}
H_0^{AB} &=& -\frac{1}{4}V'\tau_z^A\tau_z^B, \nonumber\\
H_0^{AA'}&=&\;\; \frac{1}{2}V\tau_z^A\tau_z^{A'}. 
\end{eqnarray}

The kinetic energy H$_{kin}$ consists of three terms: intra-rung hopping ($\tilde{t}$) and inter-rung hopping (t,t') (Fig.1a) along the zig-zag chain. Elimination of these processes for smalll hopping leads to an effective Hamiltonian expressed in terms of $\tau_z, \vec{\sigma}_l$. The former contains "superexchange" contributions for the pseudo spins in addition to the zeroth- order term of Eq.(1). The latter leads to a coupling between pseudo spin (i.e. charge) and spin variables. The derivation is indicated for the pseudo- spin superexchange originating in t' and $\tilde{V}$. We divide the 1-p sector of each rung with four states into ferromagnetic \{P$_0$\} and antiferromagnetic \{Q$_0$\} states characterized by projectors P$_0$ =(1+$\tau_z^A\tau_z^B$)/2 and  Q$_0$ =(1-$\tau_z^A\tau_z^B$)/2. The energy of \{Q$_0$\} states is higher by $\tilde{V}$ compared to \{P$_0$\} states. Under the assumption  $t' < \tilde{V}$ one obtains an effective 2$^{nd}$- order contribution in the 1-p subspace given by

\begin{equation}
H^{(2)}_{eff}= -P_0H_{kin}^{t'}\frac{1-P_0}{H}H_{kin}^{t'}P_0 =-\frac{2t'^2}{V'}\tau_z^A\tau_z^B.
\end{equation}

It has the same sign (ferromagnetic) as the first- order term, i.e., it favors the 3d electron of V$^{4+}$ sitting on the same side of next neighbor rungs (A,B) (Fig.1a). In a similar fashion all other kinetic terms ($\sim$ t,t') can be eliminated. Some of them connect the 1-p sector to the 2-p sector of doubly occupied rungs. Elimination of such terms give either pseudo- spin superexchange along the b- axis, or real spin exchange which, however, {\em depends} on the given pseudo spin or charge configuration. The kinetic terms $\sim \tilde{t}$ (intra-rung hopping) do not lead out of the 1-p space and therefore may be represented by the pseudospin variable $\tau^x_i$. One then obtains the effective Hamiltonian

\begin{eqnarray}
\lefteqn{H=-K'\sum_{\langle i,j\rangle}\tau_z^i\tau_z^j
   + K\sum_{\ll i,j\gg}\tau_z^i\tau_z^j -\tilde{t}\sum_{i}\tau_x^i}  \\
& &+ J'\sum_{\langle i,j\rangle}\frac{1}{32}[(1+\tau_z^i)(1-\tau_z^j)
   \vec{\sigma}_2^i\vec{\sigma}_1^j +(1-\tau_z^i)(1+\tau_z^j)
   \vec{\sigma}_1^i\vec{\sigma}_2^j] \nonumber\\
& &+J\sum_{\ll i,j\gg}\frac{1}{16}[(1+\tau_z^i)(1+\tau_z^j)
   \vec{\sigma}_2^i\vec{\sigma}_2^j +(1-\tau_z^i)(1-\tau_z^j)
   \vec{\sigma}_1^i\vec{\sigma}_1^j]. \nonumber
\end{eqnarray}

By convention $\langle i,j\rangle$ denotes nearest- neighbor A,B rungs along the [110] direction and $\ll i,j\gg$ denotes next- nearest neighbor rungs (A,A') or (B,B') which exist {\em only} along the b- direction. Under the assumption that V$<$$\tilde{V}$,U the interaction constants obtained from the above elimination procedure are given by

\begin{eqnarray}
K'&=&\frac{1}{4}V'+\frac{2t'^2}{\tilde{V}} -\frac{1}{16}J',\;\;
K =\frac{1}{2}V +\frac{t^2}{\tilde{V}}-\frac{1}{8}J, \nonumber\\
J'&=&\frac{2t'^2}{U-V'},\;\; J =\frac{2t^2}{U}.
\end{eqnarray}

Here the first two terms describe an anisotropic frustrated 2D- Ising model with ferromagnetic n.n and antiferromagnetic n.n.n interactions in the pseudo- spin variables. These terms are responsible for the charge- ordering transition. The third term describes the effect of covalency in each rung and it influences the degree as well as critical temperature of charge ordering \cite{Thal}. In the limit $\tilde{t}<V'$ it may also be eliminated and only the first two terms of the pseudospin part in Eq.(3) remain, but with the replacement K'$\rightarrow$ K'+4$\tilde{t}^2$/V' in Eq.(4). The fourth and last terms in Eq.(3) couple charge ($\tau_z$) and spin ($\frac{1}{2}\vec{\sigma}$) variables in an intricate way: The AF spin exchange $\sim$ J between n.n.n spins along b is only effective if electrons are sitting on the same side of the rungs, i.e., if the two rungs i,j are charge ordered, which means that $\tau_z^i$, $\tau_z^j$ are ferromagnetically aligned. If they occupy opposite sides, i.e., if  $\tau_z^i$, $\tau_z^j$ are antiferromagnetic, there are no virtual hopping processes to doubly occupied states possible and hence the spin exchange is zero. When the rungs are completely charge ordered along the b-axis the last term in Eq.(3) describes an AF S=$\frac{1}{2}$ spin chain along b. Thus the model of Eq.(3) has the essential ingredients to describe the physics of $\alpha$'-Na$_2$VO$_5$:\\
(1) The high temperature phase corresponds to a homogeneous MV (charge disordered) phase which is insulating because inter-rung charge fluctuations are suppressed by the U and $\tilde{V}$ interactions.
(2) the primary order parameter is due to charge order in the vanadium rungs which can be described by an anisotropic frustrated 2D Ising model for the pseudo spins.
(3) Once complete charge order has developed, linear S=$\frac{1}{2}$ spin chains along b (Fig.1b) are formed which leads to the possibility of their dimerization as {\em secondary} ordering phenomenon.\\
In the following an explicit calculation for the order parameters is given. As a simple approximation we proceed in two steps. The primary OP $\langle\tau_z\rangle$ is determined by the Ising part. In mean- field (MF) approximation the charge- ordering temperature would be determined by T$_{c1}$ $\sim$ K(Q) where K(Q)= $\pm$(z'K'-zK) (z'=2, z=4) is the Fourier transform of the interaction with the charge- ordering wave vector Q$_c$=(0,0) and Q$_c$=($\frac{\pi}{a}$,$\frac{\pi}{b}$) for 'FM'(+) and 'AF'(-) order, respectively. The former corresponds to a charge- ordered structure below T$_{c1}$ consisting of S=$\frac{1}{2}$ (V$^{4+}$) spin chains along b (dotted lines in Fig.1b). This is precisely the type of order which was originally thought to exist in $\alpha$'-NaV$_2$O$_5$ for {\em all} temperatures. The latter possibility would correspond to a charge ordering with $V^{4+}$ ions filling every second zig-zag chain completely. This Ising type model gives a very simplified picture of the real charge- order transition in an insulator where the compensation effect of long range Coulomb contributions is important to achieve a low transition temperature as in $\alpha$'-NaV$_2$O$_5$. In a rudimentary way this compensation is included in the Ising terms of Eq.(3) due to their opposite signs. The resulting reduction of the ordering temperature can roughly be incorporated by mapping this anisotropic frustrated model to the isotropic unfrustrated n.n. Ising model, i.e. by replacing K'$\rightarrow$ K$_1$ =K'-(z/z')K =K'-K/2 which would be exact in mean- field approximation. Note that the direct contribution in K$_1$ $\sim$(V'-V) is strongly reduced due to the compensation effect. Then the charge- order parameter $\langle \tau_z\rangle$ is given by Onsager's solution

\begin{equation}
\langle \tau_z\rangle =\{1-\sinh(2\beta K_1)^{-4}\}^{\frac{1}{8}}.
\end{equation}

It is assumed that K$_1$ $>$0 corresponding to a FM Ising OP describing the charge order which leads to the straight spin chains along b (Fig.1b). The AF charge order with isolated zig-zag chains is probably not favored \cite{Yaresko}. The charge- order temperature corresponding to Eq.(5) is given by T$_{c1}$=2.27K$_1$. The order parameter is shown in Fig.2 (dashed line). Its knowledge is essential to obtain the effective spin Hamiltonian from Eq.(3). thought to be responsible for the secondary ordering at T$_{c2}$ which opens the spin- excitation gap. Replacing $\tau_z$ $\rightarrow$ $\langle\tau_z\rangle$ in Eq.(3) we notice the following: The increase in $\langle\tau_z\rangle$ leads to a rapid quenching of contributions coming from the l=1 chains containing the left V atom of each rung (Fig.1b). Therefore only the intra- chain exchange contributions from l=2 chains survive. Consequently the inter- chain exchange coupling $\sim$ J' between the l=1,2 type chains also becomes ineffective below T$_{c1}$. Thus the Hamiltonian in Eq.(3) describes in a natural way how a linear spin- chain system appears below the charge ordering transition. Once it is established it may undergo a SP phase transition at T$_{c2}$ $<$ T$_{c1}$ leading to a dimerization order parameter of the linear (l=2) chains along b. Its effective 1D Hamiltonian is extracted from Eq.(3) by defining $\hat{\tau}_z$=$\tau_z$-$\langle\tau_z\rangle$ and using

\begin{equation}
\langle(1+\tau_z^i)(1+\tau_z^j)\rangle =[(1+\langle\tau_z\rangle)^2-1]
+[1+\langle\hat{\tau}_z^i\hat{\tau}_z^j\rangle].
\end{equation}

where the first term is the 1D- coherent part along b and the second term is an incoherent 2D part already present above T$_{c1}$. This separation is equivalent to a virtual- crystal approximation for the spin part. Keeping only the first coherent part the effective AF spin- chain Hamiltonian ($\vec{S}_i=\frac{1}{2}\vec{\sigma}_i$) then reads 

\begin{eqnarray}
H_{eff}^{1D}&=&J\Gamma_+(T)\sum_{\{i,j\}}\vec{S}_i\vec{S}_j, 
\nonumber\\
\Gamma_+(T)&=&\frac{1}{4}[(1+\langle\tau_z\rangle)^2-1].
\end{eqnarray}
 
Thus the charge- ordering mechanism leads to a 1D HAF with an {\em effective} temperature dependent n.n. exchange J along the b- chain (this corresponds to n.n.n exchange in the plane). The AF- spin chain may now exhibit a SP- transition if the coupling to the lattice dimerization mode, characterised by a coupling constant $\lambda$, is strong enough. The temperature dependence of the dimerization spin gap $\Delta_{SP}(T)$ is obtained from the gap equation \cite{Pytte}

\begin{eqnarray}
\frac{1}{\lambda}&=&\int_{0}^{\epsilon_0}\frac{\Gamma_+^2(T)}{[\Gamma_+^2(T)
\epsilon^2+\Delta_{SP}^2(T)]^\frac{1}{2}}\nonumber\\
& &\times\tanh[\frac{(\Gamma_+^2(T)\epsilon^2+\Delta_{SP}^2(T))^\frac{1}{2}}{2kT}]
d\epsilon.
\end{eqnarray} 
  
The cutoff is given by $\epsilon_0$= pJ with p= 1.63 \cite{Pytte}. Note that in addition to the intrinsic BCS temperature dependence there is an additional one caused by the explicit appearance of $\Gamma_{+}$(T) in the effective exchange which has the following consequences:\\
(1) because $\Gamma_{+}$(T)$\equiv$0 for T$>$T$_{c1}$ one always has T$_{c2}$ $<$T$_{c1}$ for the SP- transition.\\
(2) the temperature variation of $\Delta_{SP}(T)$ is not BCS- like due to the influence of $\Gamma_{+}$(T) and as a consequence the BCS- ratio 2$\Delta_{SP}$(0)/T$_{c2}$ will be anomalous. \\
We may however define a fictitious SP- transition temperature T$_{c0}$ through Eq.(8) which would exist if one had spin chains with a {\em constant} coupling strength J$\Gamma_+$(T) $\equiv$ J$\Gamma_+$(0) for all temperatures. It is more convenient to use T$_{c0}$ as parameter instead of $\lambda$. If T$_{c0}$ $>$ T$_{c1}$ the BCS- ratio for the SP transition at T$_{c2}$ will become anomalous. As T$_{c0}$ $\gg$ T$_{c1}$ the physical SP transition at T$_{c2}$ will be very close below the charge- order transition, i.e. (T$_{c1}$- T$_{c2}$)/T$_{c1}$ $\ll$1. These findings are clearly illustrated in Fig.2 showing the temperature dependence of both, charge ($\langle\tau_z\rangle$, dashed line) and dimerization ($\Delta_{SP}$, full line) order parameters. We identify T$_{c1}$ with the charge order temperature at 33K. Then the only parmeter is T$_{c0}$/T$_{c1}$ which is fixed to obtain the anomalous BCS- ratio of 6.44. From this we obtain a SP T$_{c2}$ of 30.4K or (T$_{c1}$- T$_{c2}$)/T$_{c1}$= 0.08. This is still too high compared to the extremely small experimental value of 0.009. However this value is determined by the details of the temperature dependence of $\langle\tau_z\rangle$ below T$_{c1}$ which is approximated by the 2D Ising model OP. In reality the charge- order transition has first- order character \cite{Koeppen} and this may push T$_{c2}$ even closer to T$_{c1}$.\\
In conclusion, the present model qualitatively can account for the crucial observations in $\alpha$'-NaV$_2$O$_5$ outlined in the beginning. It has the advantage that it gives a natural explanation for two seemingly unrelated facts: The double phase transition with T$_{c1}$ and T$_{c2}$ values close to each other and the very anomalous BCS- ratio. As a further illustration of this fact we show the dependence of this ratio as a function of the T$_{c1}$- T$_{c2}$ - splitting obtained from our model where T$_{c1}$/T$_{c0}$ is the parameter varied (insert of Fig.2). Independent of the details of the model it has become clear that charge ordering and spin (Peierls) gap formation are intricately linked in $\alpha$'-Na$_2$VO$_5$ and an important task for future investigations is therefore a precise determination of the type of charge order. 

\section*{Acknowledgement} The authors would like to thank A.N.Yaresko for useful discussions

\begin{figure}
\centerline{\psfig{figure=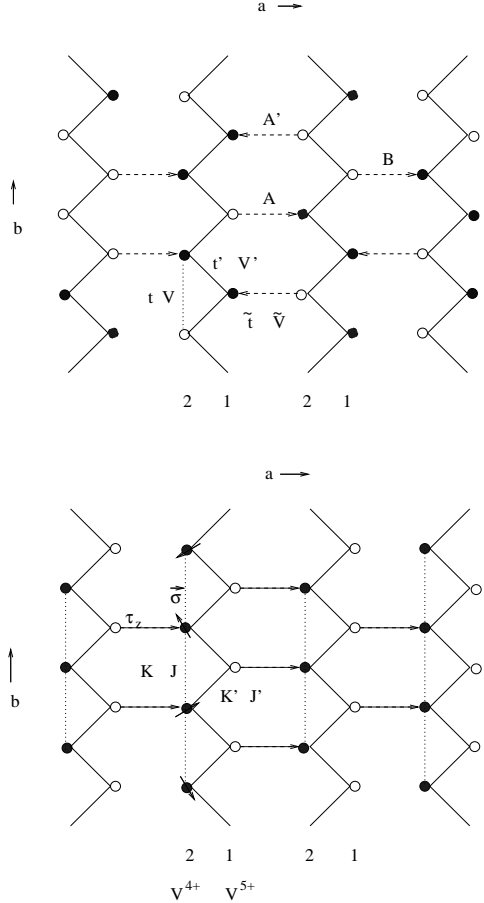,width=.5\textwidth}}
\caption{Top: Charge disordered (mixed valent) structure of $\alpha$'-NaV$_2$O$_5$ above T$_{c1}$ in the ab- plane. Each 'rung' (dashed arrows A,B,A' etc.) contains two vanadium atoms (1,2), one in the V$^{4+}$ (S=$\frac{1}{2}$) state (full circles), the other in the spinless  V$^{5+}$ state (open circles). The direction of the arrow indicates the sign of the pseudo spin $\langle\tau_z\rangle$. The configuration shown belongs to the 1-p sector for each rung. Hopping matrix elements t,t',$\tilde{t}$ and inter-site Coulomb interactions  V,V',$\tilde{V}$ are indicated. They correspond to 2$^{nd}$ (3.44 $\AA$), 1$^{st}$ (3.05 $\AA$) and 3$^{rd}$ (3.61 $\AA$) nearest neighbors. The position of O- and Na- atoms is not shown.
Bottom: charge ordered (FM-) structure of $\tau_z$ below T$_{c1}$. V$^{4+}$- ions align in linear (2- or 1- type)  S=$\frac{1}{2}$ spin ($\vec{\sigma}$)- chains along b with intra- chain spin exchange J. Inter- chain exchange J' is not active in this charge- ordered structure.}
\end{figure}

\newpage

\begin{figure}
\centerline{\psfig{figure=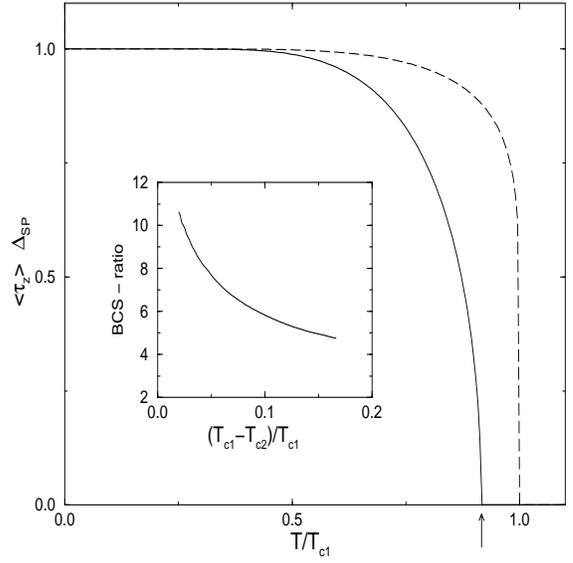,height=.5\textwidth,width=.5\textwidth,angle=-90}}
\vspace{1cm}
\caption{Temperature dependence of the {\em normalized} charge ($\langle\tau_z\rangle$, dashed line) and dimerization ($\Delta_{SP}$, full line) order parameters. Critical temperatures are T$_{c1}$= 33K and T$_{c2}$= 30.4K (arrow) respectively. A coupling parameter $\lambda$ corresponding to T$_{c0}$= 57K has been used which leads to a BCS-ratio 2$\Delta_{SP}$(0)/T$_{c2}$ =6.47. The insert shows the variation of the BCS- ratio with the normalized difference of critical temperatures.}
\end{figure} 
\end{document}